\documentclass[12pt]{iopart}
\usepackage[dvips]{graphicx}
\usepackage{subfigure}
\usepackage{iopams}
\begin{document}

\title{Mesoscopic description of reactions under anomalous diffusion: A case study}

\author{M.G.W. Schmidt}
\address{Institut f\"ur Physik, Humboldt-Universit\"at zu Berlin,
Newtonstr. 15, D-12489, Berlin, Germany} \address{Departament de
Qu\'imica F\'isica, Universitat de Barcelona, Mart\'i i Franqu\`es
1, E-08028, Barcelona, Spain}
\author{F. Sagu\'es}
\address{Departament de Qu\'imica F\'isica, Universitat de Barcelona,
Mart\'i i Franqu\`es 1, E-08028, Barcelona, Spain}
\author{I.M. Sokolov}
\address{Institut f\"ur Physik, Humboldt-Universit\"at zu Berlin,
Newtonstr. 15, D-12489, Berlin, Germany}



\begin{abstract}
Reaction-diffusion equations deliver a versatile tool for the
description of reactions in inhomogeneous systems under the
assumption that the characteristic reaction scales and the scales of
the inhomogeneities in the reactant concentrations separate. In the
present work, we discuss the possibilities of a generalization of
reaction-diffusion equations to the case of anomalous diffusion
described by continuous-time random walks with decoupled step length
and waiting time probability densities, the first being Gaussian or
L\'evy, the second one being an exponential or a power-law lacking
the first moment. We consider a special case of an irreversible or
reversible $A \rightarrow B$ conversion and show that only in the
Markovian case of an exponential waiting time distribution the
diffusion- and the reaction-term can be decoupled. In all other
cases, the properties of the reaction affect the transport operator,
so that the form of the corresponding reaction-anomalous diffusion
equations does not closely follow the form of the usual
reaction-diffusion equations.
\end{abstract}

\maketitle

\section{Introduction}

Many phenomena in systems out of equilibrium can be described using
a picture of reaction-diffusion. Examples can be found in various
disciplines, above all in chemistry but also in physics, ecology and
others. Examples from physics are the trapping and annihilation of
excitons and the electron-hole recombination in solids. In ecology,
there are e.g. the predator-pray relations. Both reaction-diffusion
with normal and anomalous diffusion have been extensively studied
over the past decades. However, for the latter, a general
theoretical framework is still absent. In this article, we discuss a
special case of the monomolecular conversion under subdiffusion and
show that the mesoscopic approach leads to equations different in
form from what could be regarded as a straightforward generalization
of the reaction-diffusion scheme.

The mesoscopic approach leading to reaction-diffusion equations is
valid if there is a strong scale separation between the typical
reaction scale and the size of the system's inhomogeneities. The
corresponding reaction-diffusion equations (for normal diffusion)
typically have the form
\begin{equation}
\frac{\partial C_i(t)}{\partial t} =K_i \Delta C_i \pm \kappa_i C_1^{n_1}
 C_2^{n_2} \cdots C_N^{n_N},
\label{RDE}
\end{equation}
which simply follows by adding a diffusion term to a classical
kinetic equation for the corresponding reaction. Here, $K_i$ denotes
the diffusivity of the component $i$, the integer powers $n_j$
correspond to the stoichiometry of the reaction, and $\kappa_i$
denotes the corresponding reaction rate.

However, many physical systems exhibit anomalous diffusion (see e.g.
\cite{MetzKlaf,Phys2day,PhysWorld} for reviews and popular
accounts), which is not adequately described by Fick's law. Many
cases of subdiffusion are successfully modeled within the
continuous-time random walk framework (CTRWs) with power-law on-site
waiting time distributions lacking the first moment. These
distributions typically have the form $w(t) \propto t^{-1-\alpha}$
with $0<\alpha <1$. Examples include, among others, dispersive
charge transport in disordered semiconductors, contaminant transport
by underground water and motion of proteins through cell membranes.
On the other hand, successful search strategies in animal motion can
be described by L\'evy walks or flights, often in combination with
broad waiting time distributions. L\'evy flights are also used as a
model for the transport on annealed polymer chains \cite{Chain,
BroGei}, which may be relevant for the gene expression
\cite{MetzlerNew}.

For anomalous diffusion, the Fickian diffusion equation is changed
for an anomalous diffusion equation involving fractional
derivatives. For subdiffusion, the equation for the concentration
$C(x,t)$ of diffusing particles reads
\begin{equation}
\frac{\partial C(x,t)}{\partial t} = K_\alpha \,_{0}D_{t}^{1-\alpha}
\Delta C(x,t), \label{subd}
\end{equation}
with the corresponding (anomalous) diffusion coefficient $K_\alpha$,
where $\,_{0}D_{t}^{\beta}$ stands for the operator of a fractional
Riemann-Liouville derivative,
\begin{equation}
\,_{a}D_{x}^{\beta}f(x)=\frac{d^n}{dx^n}\frac{1}{\Gamma(\nu)}\int_{a}^{x}\frac{f(x')}
{(x-x')^{1-\nu}}dx'
\end{equation}
with $n =[\beta]+1$ ($[x]$ stands for the whole part of the number
$x$) and $\nu = n-\beta$. For a L\'evy flight, i.e. the random walk
process with the power-law distribution of the step lengths,
$\lambda(x) \propto x^{-1 - \mu}$, the corresponding equation reads
\begin{equation}
\frac{\partial C(x,t)}{\partial t} = K_\mu  \Delta^{\mu/2} C(x,t),
\label{superd}
\end{equation}
where $ \Delta^{\mu/2}$ stands for the Riesz symmetric fractional
derivative acting on the spatial variable \cite{SZ97}. For a
''sufficiently well-behaved'' function $f(x)$ it can be expressed
through the Liouville - Weyl derivative \cite{SKM93}:
\begin{equation}
\Delta^{\mu/2}f(x)=-\frac{1}{2\cos (\pi \mu)}
\left[\,_{-\infty}D_{x}^{\mu }+\,_{x}D_{\infty}^{\mu}\right]
\end{equation}
for $\mu \neq 1/2$,
and for $\mu =1/2$ through the derivative of the Hilbert transform of $f$:
\begin{equation}
\Delta^{1/2}f(x)=-\frac{d}{dx}\frac{1}{\pi }\int_{-\infty
}^{\infty }\frac{\phi (\xi )d\xi }{x-\xi }.
\end{equation}

Reactions under anomalous diffusion were discussed by several
authors. However, most attention was payed to the description of the
elementary act of reaction on the microscopic scale \cite{micro1,
micro2, micro3, micro4, micro5}. Mesoscopic approaches were used
e.g. in \cite{Wearnes} for subdiffusion, where equations of the type
\begin{equation}
\frac{\partial C_i(\mathbf{r},t)}{\partial t} = K_{i,\alpha_i} \:_0
D_t^{1-\alpha_i} \Delta C_i(\mathbf{r},t) + f_i
\label{RDF}
\end{equation}
were postulated for different components in a multi-component
system, and in Refs. \cite{front1,front2}, where front propagation
was discussed for symmetric and asymmetric L\'evy flights,
respectively, see also Ref.\cite{Zanette} and Ref.
\cite{MetzlerNew}, where a L\'evy diffusion term was added to a
``normal'' reaction-diffusion equation to describe target search
processes on the DNA.

In what follows, we discuss the derivation of the reaction-anomalous
diffusion equations for a special case of the simple monomolecular
conversion $A \rightarrow B$ under a CTRW transport mechanism (where
our approach however differs from the one of our previous publication
Ref.\cite{ourarticle}). We consider subdiffusion, L\'evy flights and
the combination of both. Moreover, a reversible conversion $A
\rightleftharpoons B$ is also considered. As we proceed to show, the
Markovian situation of a (symmetric) L\'evy flight is described
correctly by the reaction-superdiffusion equation
\begin{equation}
\frac{\partial C_i(\mathbf{r},t)}{\partial t} = K_{\mu_i}
\Delta^{\mu_i/2} C_i(\mathbf{r},t) + f_i \label{fracsuper}
\end{equation}
with $C_i$ being $A$ or $B$ and the reaction terms $f_i=\pm \kappa
A$. On the other hand, the situation for the non-Markovian subdiffusive
transport is much more involved. The irreversible reaction can be
described by an equation for $A$ with the transport term depending
on the reaction rate, and the equation for the reversible case
cannot be casted in a form of something resembling a
reaction-diffusion equation.

The article is structured as follows: In Section \ref{educt} we
derive the equation for the time evolution of the educt
concentration $A$ in an irreversible reaction. The behavior of the
product concentration $B$ is discussed in section \ref{product}.
Section \ref{reversible} is devoted to a mesoscopic approach to
reversible conversions.

\section{The educt concentration in the irreversible conversion $A \rightarrow B$}
\label{educt}

In what follows, we consider the situation where $A$-particles are
converted into $B$ at a constant conversion rate $\kappa$
independent on their position. Thus, the survival probability of a
single $A$-particle in the time interval $[t',t]$ is
$\Phi_{A}(t,t')=\Phi_{A}(t-t')=\exp[-\kappa(t-t')]$. We will use
one-dimensional notation in the following, the generalization to
higher dimensions is straightforward. An example for this situation
is the decay of a radioactive isotope in the groundwater, where the
reaction and the transport mechanism are fully decoupled. We are
interested in the mathematical description of the situation, where
the transport is given by a decoupled CTRW process with given step
length and waiting time distribution. Our derivation of
reaction-anomalous diffusion equations is parallel to the derivation
of the pure anomalous diffusion equations in \cite{MetzKlaf}.

We can put down an equation for the probability density function
(pdf) of the positions $x$ of the particles, which have just made a
jump at time $t$:
\begin{eqnarray}
\eta_{A} (x,t)=\int_{-\infty}^{\infty}\int _{0}^{t}\eta_{A}(x',t')e^{-\kappa(t-t')}
\psi(x-x',t-t')dx'dt'+A(x,0)\delta(t).
\label{ansatzetaA}
\end{eqnarray}
Here, $\psi(x,t)$ is the jump pdf given by the probability density
in space and time to make a jump of length $x$ at time $t$ after the
last jump. The meaning of the equation is that for whatever $t>0$ an
$A$-walker that has just arrived at $x$ has come there from some
other site, where it had survived as $A$ during the whole waiting
time. The second term corresponds to the initial condition that at
time $t=0$ all particles are assigned a new waiting time. Here, we
have additionally assumed that the jump length distribution does not
depend on the position of the walker and that the waiting time pdf
is constant in time and space. Furthermore, $\psi(x-x',t-t')$ is
assumed to be decoupled $\psi(x-x',t-t')=\lambda(x-x')w(t-t')$.

In order to get the equation of motion for the $A$-particles, i.e.
for the concentration $A(x,t)$, we connect it to $\eta_{A}(x,t)$
over
\begin{eqnarray}
A(x,t)=\int_{0}^{t}dt'\eta_{A}(x,t')e^{-\kappa(t-t')}\Psi(t-t'),
\label{ansatzA}
\end{eqnarray}
where $\Psi(t-t')$ is the probability to stay at site $x$ for a time
$(t-t')$ after a jump. It is given by
\begin{eqnarray}
\Psi(t)=1-\int_{0}^{t}dt'w(t').
\label{Psi}
\end{eqnarray}
Both Eqs.(\ref{ansatzetaA}) and (\ref{ansatzA}) contain convolution
integrals and can be solved by Fourier-Laplace transform. Using the
shift theorem for the Laplace transform, we get
\begin{eqnarray}
\hat{\tilde{A}}(k,u)=\frac{[1-\tilde{w}(u+\kappa)]\hat{A}(k,0)}
{(u+\kappa)[1-\hat{\tilde{\psi}}(k,u+\kappa)]}. \label{Montroll}
\end{eqnarray}
Before we can return to the space- and time-domain, we have to
specify the jump pdfs. We are interested in the continuum limit of
the equations corresponding to large scales and long times, i.e. to
$(k,u) \rightarrow (0,0)$. A characteristic function of a Gaussian
jump length pdf with variance $2 \sigma^2$ will then be approximated
by $\hat{\lambda}(k) \simeq 1-k^{2}\sigma^{2}$. A characteristic
function of a broad L\'evy distribution,
$\hat{\lambda}(k)=\exp(-\sigma^{\mu}|k|^{\mu})$, can be approximated
through $\hat{\lambda}(k) \simeq 1-\sigma^{\mu}|k|^{\mu}$. For a
broad waiting time pdf of a Pareto (power-law) type, $w(t) \simeq \alpha
\tau^{\alpha} t^{-1-\alpha}$, one infers the following asymptotics
in Laplace space using a Tauberian theorem, $\tilde{w}(u) \simeq
1-\Gamma(1-\alpha) u^{\alpha}\tau^{\alpha}$. For the Markovian case,
as exemplified by the exponential waiting time pdf,
$w(t)=\tau^{-1}\exp(-t/\tau)$, one has $\tilde{w}(u) \simeq 1-u\tau$
in the continuum limit, which corresponds to $\alpha =1$.
Eq.(\ref{Montroll}) then can be rewritten in the following form
\begin{equation}
u \hat{\tilde{A}}(k,u) - \hat{A}(k,0) = -(u+\kappa)^{1-\alpha}
\frac{\sigma^\mu} {\Gamma(1-\alpha)\tau^\alpha} \left| k \right|^\mu
\hat{\tilde{A}}(k,u) - \kappa \hat{\tilde{A}}(k,u) \label{FurLa}
\end{equation}
simplifying the inverse transforms. For the inverse Fourier
transformation, we use $\mathcal{F}^{-1}\{-k^{2}\hat{f}(k)\}=\Delta
f(x)$, and
$\mathcal{F}^{-1}\{-|k|^{\mu}\hat{f}(k)\}=\Delta^{\mu/2}f(x)$.
Moreover, we introduce the notation $K_{\mu
,\alpha}=\sigma^\mu[\tau^\alpha \Gamma(1-\alpha)]^{-1}$ for what
later will be identified as the generalized diffusion coefficient.
The inverse Laplace transform of the left hand side of the equation is
simply the first time derivative, since $\mathcal{L}^{-1} \{u
\hat{g}(u) - g(0) \}= dg(t)/dt$.

We first combine the Gaussian and L\'evy distributed jump length pdf
with an exponential waiting time pdf. In this case, the pre-factor
of $ \hat{\tilde{A}}(k,u)$ in the first term on the right side of
the equation does not depend on $u$. After inverse transforming the
equation, it becomes a time-independent operator acting on the
concentration as a function of the coordinates. For a Gaussian jump
length distribution, our equation (\ref{FurLa}) now reads
\begin{eqnarray}
\frac{\partial A(x,t)}{\partial t}=K_{2,1}\Delta A(x,t) - \kappa A(x,t),
\end{eqnarray}
 and for L\'evy flights,
\begin{eqnarray}
\frac{\partial A(x,t)}{\partial t}=K_{\mu, 1}\Delta^{\mu/2} A(x,t) -
\kappa A(x,t).
\end{eqnarray}
Hence, the separation of the transport- and the reaction-term is
perfectly exact. For L\'evy flights, the Laplace operator is just
changed for the Riesz-Weyl fractional derivative.

Now, we turn to subdiffusion and consider a Gaussian distribution of
the step lengths ($\mu = 2$) combined with a broad waiting time pdf
of a Pareto type with $0 <\alpha < 1$. From Eq.(\ref{FurLa}) we then
get
\begin{equation}
\frac{\partial A(x,t) }{\partial t}= K_{2,\alpha} \mathcal{T}_{t}(1-\alpha,\kappa)
\Delta A(x,t)-\kappa A(x,t),
\label{Aconc}
\end{equation}
with the transport operator $\mathcal{T}_{t}(1-\alpha,\kappa)
\Delta$, which is now time-dependent,
\begin{eqnarray}
\mathcal{T}_{t}(1-\alpha,\kappa)f(t)=\frac{1}{\Gamma(\alpha)}&\left(
\frac{d}{dt} \int_{0}^{t}\frac{e^{-\kappa(t-t^{\prime
})}}{(t-t^{\prime })^{1-\alpha }} f(t^{\prime})dt^{\prime }\right.
\nonumber \\
&\left. +\kappa \int_{0}^{t}\frac{e^{-\kappa (t-t^{\prime
})}}{(t-t^{\prime })^{1-\alpha }}f(t^{\prime })dt^{\prime }\right).
\label{operatorT}
\end{eqnarray}
Its form follows from the shift theorem for the Laplace transform.
We see that the reaction parameter enters the transport-term, and
the transport operator $\mathcal{T}_{t}(1-\alpha,\kappa)$ reduces to
a fractional derivative only for $\kappa=0$. Using the Laplace
transform property of the Riemann-Liouville fractional derivative,
$\mathcal{L}^{-1}\{u^{-\alpha}\tilde{f}(u)\}=\,_{0}D_{t}^{-\alpha}f(t)$
for $\alpha>0$, and using the shift theorem, the temporal part of a
transport operator (in its action on the arbitrary function of time
$f(t)$) can be transformed into a form \cite{Henryprivate}
\begin{equation}
\mathcal{T}_t (1-\alpha, \kappa) f(t)= \exp(-\kappa
t)\,_{0}D_{t}^{1-\alpha}
 \{ \exp(\kappa t) f(t)\}.
\end{equation}
One can also easily formulate the equations for the combination of
Pareto waiting times and L\'evy jumps being of the form
\begin{equation}
\frac{\partial A(x,t) }{\partial t}= K_{\mu ,\alpha}
\mathcal{T}_{t}(1-\alpha,\kappa) \Delta^{\mu/2} A(x,t)-\kappa
A(x,t),
\end{equation}
with $\Delta^{\mu/2}$ denoting the symmetrized (Riesz-Weyl) spatial
fractional derivative.

By the way, as shown in \cite{Metzler2}, an external force field can
be included in the model over an asymmetric jump length distribution
leading to a fractional Fokker-Planck equation with the time
fractional operator changed for our operator $\mathcal{T}_t
(1-\alpha, \kappa)$ and with an additional reaction term.

\section{Equations for the product concentration}
\label{product}

Let us turn to the equation for the concentration of the
$B$-particles. One can distinguish two different cases: (i) Either a
$B$-particle takes over the waiting time of the $A$-particle that it
was converted from, or (ii) we assign it a new waiting time when it
is produced. The former means that the conversion is just a
relabeling from the standpoint of diffusion and that conversion and
transport are totally independent. The latter is appropriate when
$A$- and $B$-particles have different diffusive properties, e.g.
when they are trapped by different kinds of molecules. Then,
transport and conversion are partly coupled.
\\
\\
\textbf{(i)} The first case corresponds to the following approach
\begin{eqnarray}
\eta_{B}(x,t)=\int_{-\infty}^{\infty}dx'\int _{0}^{t}dt'&\Bigg\{
\left[\eta_{B}(x',t')+\eta_{A}(x',t')\left(1-e^{-\kappa(t-t')}\right)\right]\times
\nonumber \\
&\times \psi(x-x',t-t')\Bigg\} +B(x,0)\delta(t), \label{ansatzetaB}
\end{eqnarray}
which expresses the fact that a $B$-particle that has just landed at
$x$ at time $t$ has come from a site $x'$ at prior time $t'$, where
it had either come already as a $B$-particle or where it had been
converted from an $A$-particle. For the concentration of
$B$-particles, we have
\begin{eqnarray}
B(x,t)=\int_{0}^{t}dt'\left[\eta_{B}(x,t')+\eta_{A}(x,t')
\left(1-e^{-\kappa(t-t')}\right)\right]\Psi(t-t'), \label{ansatzB}
\end{eqnarray}
with $\Psi(t)$ from Eq.(\ref{Psi}). Now, a $B$-particle that is at
site $x$ at time $t$ has come there at a prior time $t'$ either
already as a $B$-particle or as an $A$-particle and has been
converted in $(t-t')$. Eq.(\ref{ansatzB}) can also be solved using
Fourier-Laplace transform and Eqs.(\ref{ansatzetaA}), (\ref{Aconc})
and (\ref{ansatzetaB}). First, we get
\begin{eqnarray}
\hat{\tilde{B}}(k,u)+\hat{\tilde{A}}(k,u)=\frac{\hat{B}(k,0)+\hat{A}(k,0)}
{1-\hat{\tilde{\psi}}(k,u)}\frac{1-\tilde{w}(u)}{u},
\end{eqnarray}
which is essentially the Fourier-Laplace transformed subdiffusion
equation for the sum of the the concentrations
$C(x,t)=A(x,t)+B(x,t)$. This is due to the fact that we have assumed
a complete independence of the transport and the conversion and can
already be seen adding the two approaches Eqs.(\ref{ansatzetaA}) and
(\ref{ansatzetaB}). Using the corresponding solutions for the
concentration of $A$-particles, for a Poissonian waiting time pdf,
one infers an equation of the form (\ref{RDE}) or (\ref{fracsuper}).
For a power-law waiting time pdf and the initial conditions
$A(x,0)=\delta(x)$, $B(x)=0$ we get
\begin{eqnarray}
\frac{\partial B(x,t)}{\partial
t}=&K_{2,\alpha}\,_{0}D_{t}^{1-\alpha}
\Delta B(x,t)+\kappa A(x,t)+\nonumber \\
&+K_{2,\alpha} \left[ \,_{0}D_{t}^{1-\alpha} -
\mathcal{T}_{t}(1-\alpha,\kappa)\right] \Delta A(x,t).
\label{EqProd}
\end{eqnarray}
The change of the concentration of the $B$-particles depends on the
concentration of the $A$-particles at all previous times. This is
due to the fact that the $B$-particles are already ``aged'' when
produced and have a memory for the last jump they have made as an
$A$-particle because of the non-Markovian nature of the waiting time
pdf. As mentioned above, the combination with a L\'evy distributed
jump length pdf leads to the same result as Eq.(\ref{EqProd}) with
the Laplace operator just changed for its fractional generalization.

\begin{figure}
\centering
\includegraphics[width=0.48\textwidth]{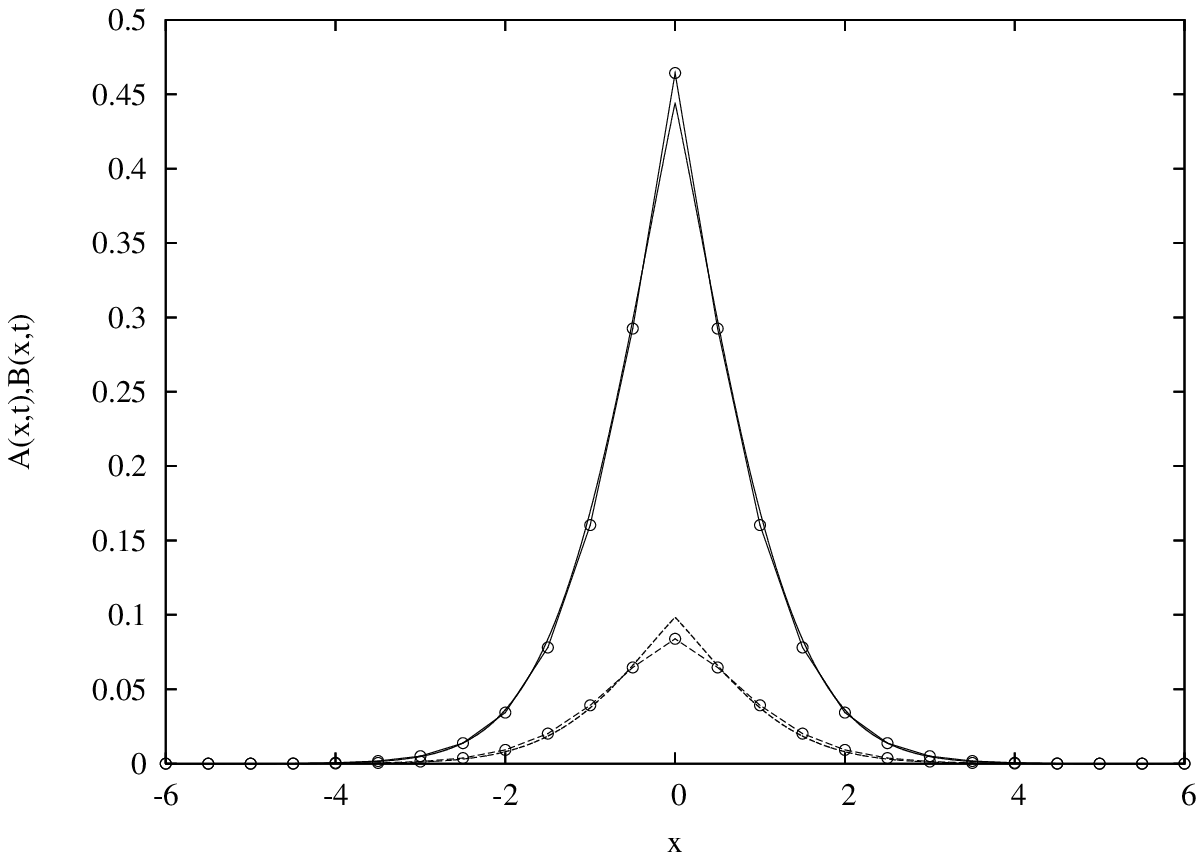}
\includegraphics[width=0.48\textwidth]{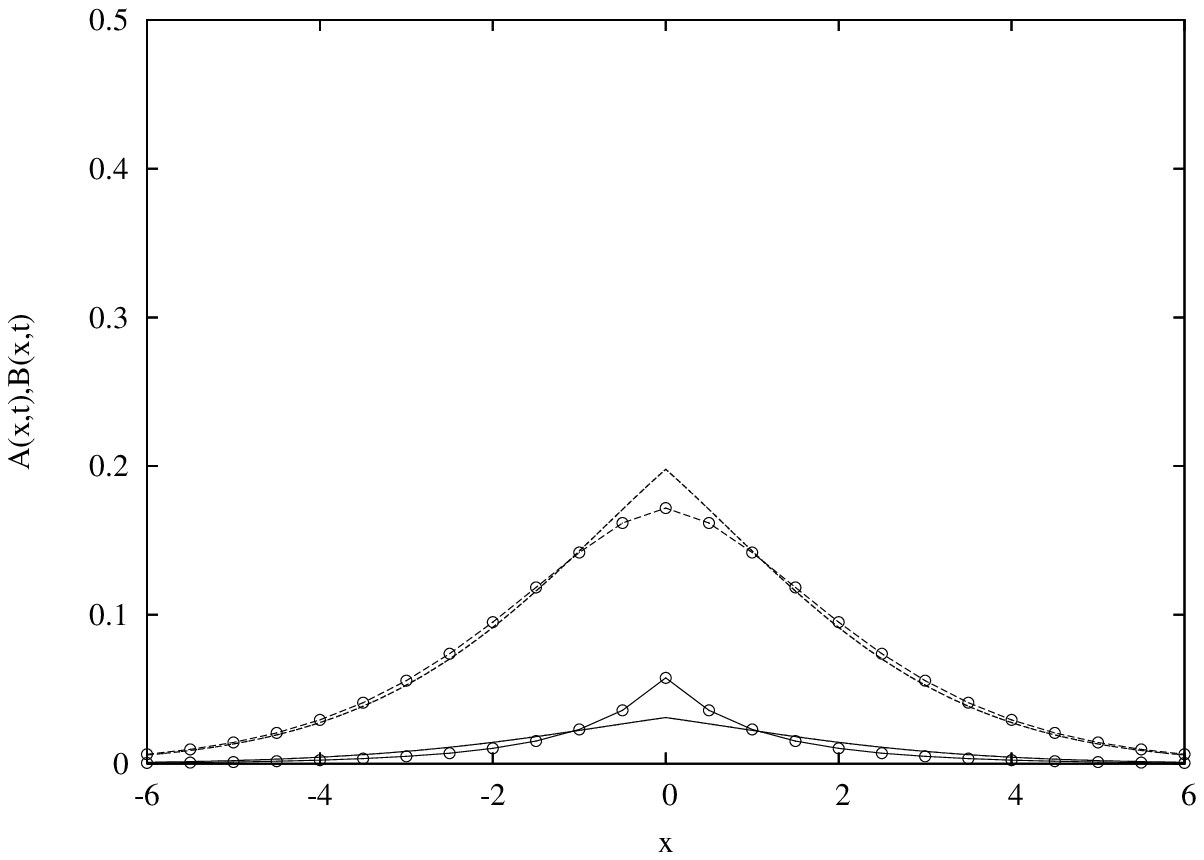}
\caption{\footnotesize Shown are the the concentrations of
$A$-particles (solid lines) and $B$-particles (dashed lines) for
subdiffusion with conversion. The correct results (solution of
Eqs.(\ref{Aconc}) and (\ref{EqProd})) are shown without dots. They
are compared to the solutions of the decoupled equations (\ref{RDF})
shown with dots. The parameters are: $\alpha=0.75$, $\kappa=0.001$,
$K_{\alpha}^{2}\simeq 7.76 \cdot 10^{-3}$. The times shown are
$t=200$ (left) and $t=2000$ (right).} \label{comparecorrect}
\end{figure}

In Fig.\ref{comparecorrect} we compare the correct solutions, i.e.
the solutions of Eqs.(\ref{Aconc}) and (\ref{EqProd}), with the
solutions of the special cases of Eq.(\ref{RDF}) for the conversion.
We note an even qualitative difference, so the latter justified only
by analogy to the normal diffusion case cannot be used as an
approximation of the exact equations.
In order to get these results, we did not actually have to solve
Eqs.(\ref{Aconc}) and (\ref{EqProd}) because we could specify the
solution from the fact that $C(x,t)$, the sum of the concentrations
of $A$- and $B$-particles, fulfills a pure subdiffusion equation.
For the conversion reaction with the reaction independent on the
transport, the concentrations are just given by
$A(x,t)=C(x,t) \exp(-\kappa t)$, $B(x,t)=C(x,t)[1- \exp(-\kappa t)]$,
namely by the product of the overall particle concentration and the
survival probability or the conversion probability, respectively.
The solution of the pure subdiffusion equation for $C(x,t)$ and the
initial condition, $C(x,0)=\delta(x)$ is known. It is the Fox's
H-function,
\begin{eqnarray}
C(x,t)=\frac{1}{4K_{2,\alpha}t^{\alpha}}H^{1,0}_{1,1}\left[\frac{|x|}
{\sqrt{K_{2,\alpha}t^{\alpha}}}\left|\begin{array}{c}(1-\alpha/2,\alpha/2)\\
(0,1) \end{array}\right.\right].
\end{eqnarray}
The Fox's H-function can be calculated using a series expansion
\cite{MetzKlaf}. The equations of the form Eq.(\ref{RDF}) were
solved using a modification of a numerical scheme presented recently
by Yuste \textit{et al} \cite{Yuste}. The scheme is a combination of
a forward-time-centered-space discretization and the
Gr\"unwald-Letnikov form of the fractional derivative.
\\
\\
\textbf{(ii)} Let us now consider the second case and assume that
$B$-particles are assigned a new waiting time at production. Here,
we expect to get a decoupled equation of the form (\ref{RDF})
because the past as an $A$-particle is ``forgotten''. We have to
start from
\begin{eqnarray}
\eta_{B}(x,t)=\int_{-\infty}^{\infty}\int
_{0}^{t}&\eta_{B}(x',t')\psi(x-x',t-t')dx'dt'+\nonumber \\
&+\kappa A(x,t)+B(x,0)\delta(t),
\end{eqnarray}
and
\begin{eqnarray}
B(x,t)=\int_{0}^{\infty}\eta_{B}(x,t')\Psi(t-t')dt'.
\end{eqnarray}
This leads first to
\begin{eqnarray}
\hat{\tilde{B}}(k,u)= \frac{1-\tilde{w}(u)}{u}\frac{\kappa
\tilde{\hat{A}}(k,u) + \hat{B}(k,0)}{1-\hat{\tilde{\psi}}(k,u)},
\end{eqnarray}
and then with a Gaussian jump length pdf and the same initial
conditions as above to
\begin{eqnarray}
\frac{\partial B(x,t)}{\partial
t}=K_{2,\alpha_{B}}\,_{0}D_{t}^{1-\alpha_{B}}\Delta B(x,t) + \kappa
A(x,t),
\end{eqnarray}
the expected decoupled equation. We have denoted the diffusion
exponent as $\alpha_{B}$ in order to emphasize that it is possibly
different from the one for the $A$-particles. By the way, instead of
the reaction-term $\kappa A$ of the conversion we could have an
arbitrary reaction term that does not depend on the product
concentration.

\section{Reversible $A \rightleftharpoons B$ reaction}
\label{reversible}

\begin{figure}
\centering
\subfigure{\includegraphics[width=0.48\textwidth]{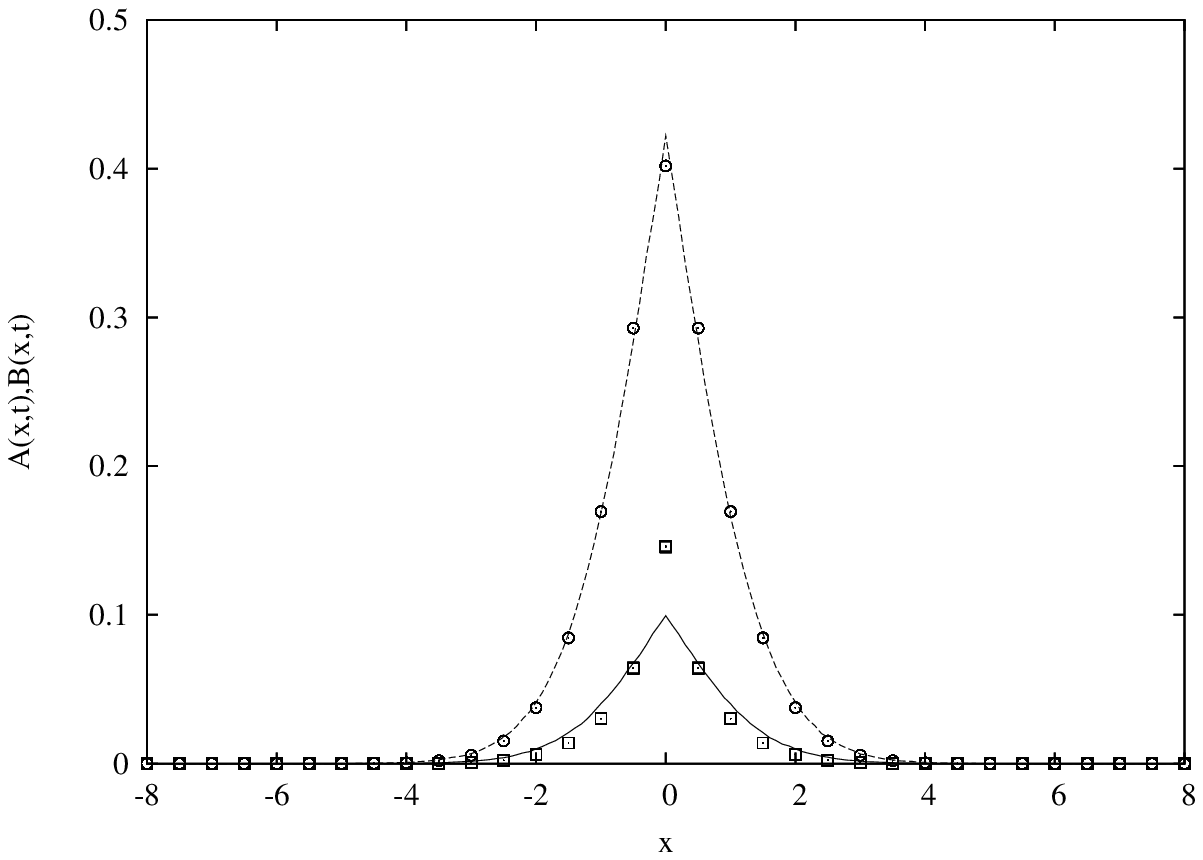}}
\subfigure{\includegraphics[width=0.48\textwidth]{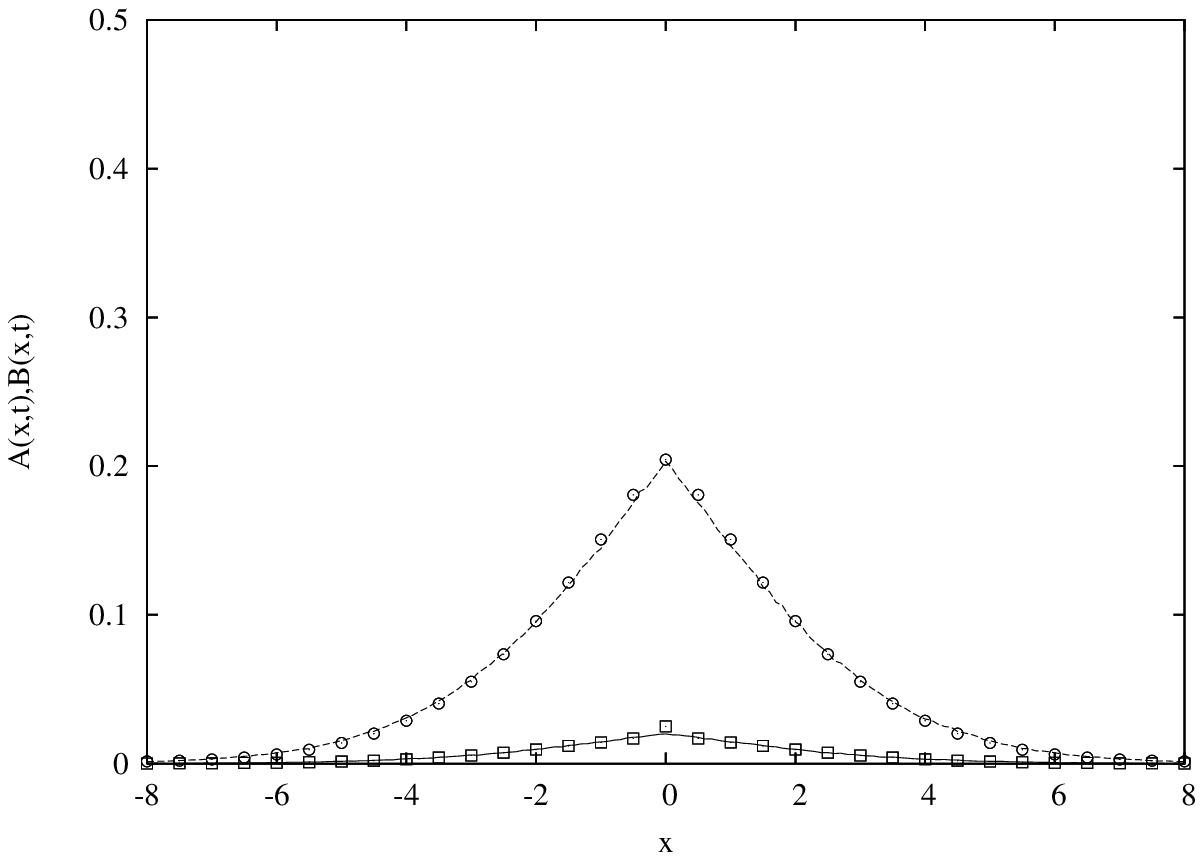}}
\caption{\footnotesize Shown is the result of the simulation:
A-particles (solid line) and B-particles (dashed line), and  the
numerical solution of the decoupled equations, Eq.(\ref{RDF}) or
Eq.(\ref{decoupled_FL}): A-particles (squares) and B-particles
(dots). The parameters are: $\alpha=0.75$, $K_{\alpha}=0.0138$,
$\kappa_{1}=0.01$, $\kappa_{2}=0.001$. The times shown are $t=200$
(left) and $t=2000$ (right).} \label{simulationgraph}
\end{figure}

Now, we turn to the case of a reversible conversion. We assume that
no new waiting time is assigned when a particle is converted. We
denote the forward reaction rate by $\kappa_1$ and the backward rate
by $\kappa_2$. Then we have to start from
\begin{eqnarray}
\eta_{A} (x,t)=\int_{-\infty}^{\infty}\int
_{0}^{t}&\left\{\left[\eta_{A}(x',t')
e^{-\kappa_1(t-t')}+\eta_B(x',t')\left(1-e^{-\kappa_2(t-t')}\right)\right]
\times \right. \nonumber \\
& \times \psi(x-x',t-t')dx'dt'\Big\} +A(x,0)\delta(t).
\end{eqnarray}
An $A$-walker that arrives at $x$ at time $t$ has come from another
site $x'$ at a prior time $t'$, where it had come already as an
$A$-particle and was not converted, or where it had come as a
$B$-particle and was converted. For the concentration, we have
\begin{eqnarray}
A(x,t)=&\int_{0}^{t}dt'\eta_{A}(x,t')e^{-\kappa_1(t-t')}\Psi(t-t')+
\nonumber \\
&+\int_{0}^{t}dt'\eta_B(x,t')\left(1-e^{-\kappa_2(t-t')}\right)\Psi(t-t').
\end{eqnarray}
An $A$-particle at site $x$ at time $t$ has come to this site
already as an $A$ at time $t'$ and has not been converted and moved
since, or it has come there as a $B$-particle, was converted and has
not moved in the mean-time. Because of the ``symmetry'' of the
reaction, the equations for the $B$-particles can be directly
inferred from the equations for the $A$-particles. We can still
perform Fourier-Laplace transform.  Using a Gaussian jump length
pdf, an inverse power-law waiting time pdf, and the initial
conditions $A(x,0)=\delta(x),\; B(x,0)=0$, we find for the
$A$-particles
\begin{eqnarray}
\hat{\tilde{A}}(u,k)=&\frac{\qquad\qquad\;\;\;\;\;\; K_{2,\alpha}
k^2+[(u+\kappa_1)-(u+\kappa_1)^{1-\alpha}u^\alpha]\times} {[K_{2,\alpha}
k^2+(u+\kappa_2)^\alpha][K_{2,\alpha}
k^2(u+\kappa_1)^{1-\alpha}+(u+\kappa_1)]+} \cdots \nonumber
\\
&\cdots\frac{\times[u^{\alpha-1}-(u+\kappa_2)^{\alpha-1}]+(u+\kappa_2)^\alpha
\qquad\qquad}
{+[u^\alpha(u+\kappa_1)^{1-\alpha}-(u+\kappa_1)][(u+\kappa_2)^\alpha-u^\alpha]}.
\end{eqnarray}
However, after the inverse Fourier-Laplace transform, the equation of motion
does not take any simple form, let alone the form of a reaction-diffusion equation. 
The decoupled scheme,
Eq.(\ref{RDF}), corresponds to a different equation,
\begin{eqnarray}
\hat{\tilde{A}}(u,k)=\frac{K_{2,\alpha}k^2 + \kappa_2 u^{\alpha-1}+
u^\alpha}{[K_{2,\alpha}k^2 + u^\alpha][K_{2,\alpha}k^2 u^{1-\alpha}
+u+\kappa_1 +\kappa_2]}. \label{decoupled_FL}
\end{eqnarray}
We have made some simple simulations in order to test how the
decoupled equations perform for this case. For the conversion, the
random walkers are independent, and we can simply repeat the random
walk procedure many times ($10^6$ times). We used the power law
waiting time pdf with a cut-off at small times guaranteeing the normalization, 
$w(t)=\alpha \tau^{\alpha} t^{-1-\alpha}$ for $t>\tau$ and $w(t)=0$ otherwise. The
conversion is independent of the jumps and takes place with a
constant probability $P_{A,(B)}=[1-\exp(-\kappa_{1,(2)}\Delta t)]$ in
each time-step of length $\Delta t$. We can see in
Fig.\ref{simulationgraph} that the coincidence with the correct
result is somewhat better for large times than in the case of an
irreversible conversion.

\section{Conclusions}

We discussed generalizations of the reaction-diffusion scheme to the
case of anomalous diffusion for a special case of a simple
conversion reaction $A \rightarrow B$ or $A \rightleftharpoons B$.
Although the reaction and the particle transport were assumed to be
independent, the reaction-term and the transport-term do not
separate in the case of subdiffusion. This means that the transport
operator in the corresponding equations depends on the properties of
the reaction. The simple equations with separated reaction- and
diffusion-terms are not exact. Comparing the exact solution with the
equations with decoupled reaction- and diffusion-terms shows that
the latter deliver a rather poor approximation for the case of an
irreversible reaction and perform somewhat better in the reversible
case.

\end{document}